\documentclass{article}

\usepackage[latin1]{inputenc}
\usepackage{import} 

\usepackage[pdftex]{graphicx}
\usepackage{color}
\usepackage{transparent} 

\graphicspath{{fig/}}

\title{CLARO-CMOS, a very low power ASIC for fast photon counting with pixellated photodetectors}

\author{Paolo Carniti$^{a,b}$, Marcello De Matteis$^{c}$, Andrea Giachero$^{a,b}$,\\
Claudio Gotti$^{a,d}$\thanks{Corresponding author}, Matteo Maino$^{a,b}$, Gianluigi Pessina$^{a,b}$\\
\\
\small
\llap{$^a$}INFN, Sezione di Milano Bicocca,
\small
Piazza della Scienza 3, 20126, Milano, Italy\\
\small
\llap{$^b$}Dipartimento di Fisica G. Occhialini, Universit\`a degli Studi di Milano Bicocca,\\
\small
Piazza della Scienza 3, 20126, Milano, Italy\\
\small
\llap{$^c$}Dipartimento di Ingegneria dell'Innovazione, Universit\`a del Salento,\\
\small
Via Per Monteroni, 73100, Lecce, Italy\\
\small
\llap{$^d$}Dipartimento di Elettronica e Telecomunicazioni, Universit\`a degli Studi di Firenze,\\
\small
via S. Marta 3, 50139, Firenze, Italy\\
\small
 E-mail: claudio.gotti@mib.infn.it}

\begin{document}

\date{}

\maketitle

\begin{abstract}
The \mbox{CLARO-CMOS} is an application specific integrated circuit (ASIC) designed for fast photon counting with pixellated photodetectors such as multi-anode photomultiplier tubes (Ma-PMT), micro-channel plates (MCP), and silicon photomultipliers (SiPM).
The first prototype has four channels, each with a charge sensitive amplifier with settable gain and a discriminator with settable threshold, providing fast hit information for each channel independently.
The design was realized in a long-established, stable and inexpensive \mbox{0.35 $\mu$m} CMOS technology, and provides outstanding performance in terms of speed and power dissipation. The prototype consumes less than \mbox{1 mW} per channel at low rate, and less than \mbox{2 mW} at an event rate of \mbox{10 MHz} per channel. The recovery time after each pulse is less than \mbox{25 ns} for input signals within a factor of 10 above threshold.
Input referred RMS noise is about \mbox{7.7 ke$^-$} (\mbox{1.2 fC}) with an input capacitance of \mbox{3.3 pF}. Thanks to the low noise and high speed, a timing resolution down to \mbox{10 ps} RMS was measured for typical photomultiplier signals of a few million electrons, corresponding to the single photon response for these detectors.
\end{abstract}

\noindent \textbf{Keywords:} Pixelated detectors and associated VLSI electronics; Cherenkov detectors; Instrumentation and methods for time-of-flight (TOF) spectroscopy; Analogue electronic circuits; Front-end electronics for detector readout; VLSI circuits

\section{Introduction}

The fast counting of photons down to the single photon level is a basic requirement shared among several applications, ranging from particle identification in fundamental physics to imaging of biological processes in nuclear medicine.
In many cases the applications require pixellated photodetectors with pixel size of the order of a few squared millimeters, often placed side by side to increase the total photosensitive area.
The total number of pixels can be very large, ranging up to the order of 10$^5$.
The case of ring imaging Cherenkov (RICH) detectors is one of the most demanding. In this case, the photodetectors are usually arranged to form planes of up to a few squared meters, ideally with no dead space between pixels.

Among the photodetectors which may be employed, multi-anode photomultiplier tubes (Ma-PMT), thanks to the negligible dark count rate, are most often the baseline for RICH detectors. New time of flight (TOF) detector designs often employ light sensors with superior time resolution, such as microchannel plates (MCP). Scintillator-based detectors usually generate a larger number of photons per event, and can thus take advantage of light detectors with a higher dark count rate, but lower cost, such as silicon photomultipliers (SiPM). From the point of view of the readout electronics, the signals from these photodetectors have very similar characteristics, and the same readout circuits can be used with minor adjustments. The typical photomultiplier gain is of the order of 10$^6$, and the expected pixel capacitance is of the order of a few pF. The charge collection time is small, of the order of one nanosecond. Other kind of photosensors exist which require different design solutions for the readout electronics, but are not considered here.

The main challenges in the realization of the electronic readout of such systems stem from the large number and close packing of the readout channels.
This requires a low power dissipation to minimize cooling issues. Other frequent requirements are the sustainability of high count rates or the allowance of precise timing measurements. These call for wide bandwidth, which is in contrast with low power dissipation.
Wide bandwidth also requires the minimization of the capacitance between pixels, which can be a major source of crosstalk.
To mimimize capacitance the front end electronics must be as close as possible to the photosensors, which also helps in minimizing noise. But this poses design issues which go back to power dissipation and cooling.
These trade-offs need to be tuned to the specificities of each application.

\begin{figure}[t]
\centering
\includegraphics[width=0.5\linewidth]{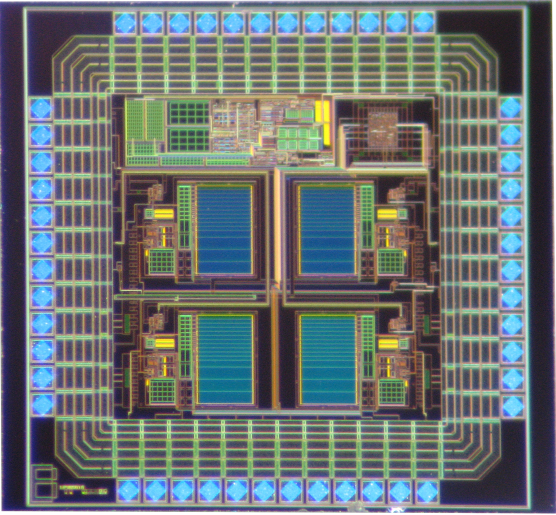}
\caption{A photograph of the 4-channel CLARO-CMOS prototype.}
\label{clarophoto}
\end{figure}

Several application specific integrated circuits (ASIC) for photodetector readout are already available, covering a wide range of applications \cite{nino, maroc, peta, spiroc, basic, vata, rapsodi}.
For instance, an ASIC suitable for timing measurements with a resolution of \mbox{20 ps} RMS is the NINO \cite{nino}, designed in IBM \mbox{0.25 $\mu$m} CMOS technology, with a power consumption of \mbox{27 mW} per channel. On the other side, an ASIC for fast photon counting with a lower power consumption is the MAROC \cite{maroc}, designed in AMS \mbox{0.35 $\mu$m} SiGe-BiCMOS technology, which consumes about \mbox{5 mW} per channel but was not designed for precise timing measurements.

The technological advances driven by the field of digital electronics and of commercial portable communication devices, which also require wide bandwidth at low power, can result in significant improvements in the field of fast photodetector readout.
Nevertheless careful design in a rather aged (and inexpensive) purely CMOS technology, such as the \mbox{0.35 $\mu$m} from AMS, can still yield excellent results at the cutting edge of timing performance and low power.
This is the aim followed in the design of the CLARO-CMOS, the first prototype of an ASIC for photodetector readout presented in this paper.
Figure \ref{clarophoto} shows a photograph the 4-channel ASIC. The die area is \mbox{2 $\times$ 2 mm$^2$}.

The radiation hardness of the technology adopted is expected to be adequate for most accelerator and space environments \cite{radhard1, radhard2}. However the effects of radiation on the circuit performance depend also on the design and layout of a given device. The radiation hardness of the CLARO-CMOS prototype will be measured in the near future, but is not considered in this paper.

\section{Design of the prototype}

\begin{figure}[t]
\centering
 \def\svgwidth{\linewidth}
 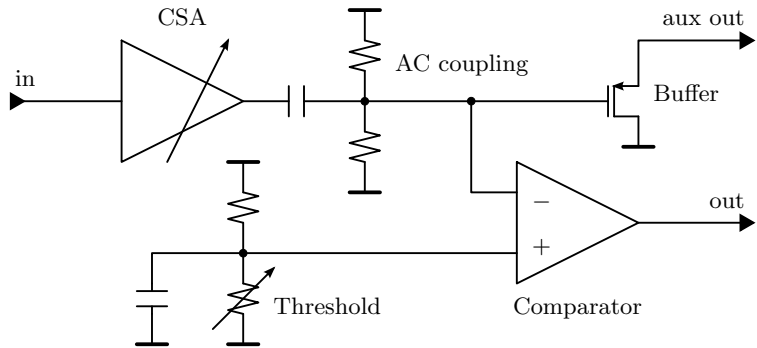
\caption{The block diagram of a CLARO-CMOS channel.}
\label{schchan}
\end{figure}

Figure \ref{schchan} shows the block diagram of a channel of the CLARO-CMOS. The ASIC is designed for operation between a positive \mbox{2.5 V} supply rail and ground.
The charge sensitive amplifier (CSA) converts the input current pulse into a voltage signal, which is AC coupled to a PMOS follower and to a discriminator (a voltage comparator). The threshold of the discriminator is set by the programmable static voltage at the non-inverting input of the comparator. The schematics of the charge sensitive amplifier and of the comparator will be described in detail in the following.

As will become clear later, the DC voltage at the output of the CSA is close to the positive rail and its value is not stable against temperature variations. For these reasons the AC coupling shown in figure \ref{schchan} was introduced.
In this way the DC voltage at the inverting input of the comparator is held at half-way between the positive rail and ground, and is independent of temperature.
The AC coupling time constant is \mbox{55 ns}. Since, as will be shown, the signals at the ouput of the CSA are very fast, no noticeable baseline shift is caused by the AC coupling unless the rate is larger than about \mbox{10 MHz}.

The auxiliary output buffer realized with a small area PMOS follower is primarily used for debugging purposes: it allows to measure the signals at the inverting input of the discriminator without loading the output of the CSA. It needs to be biased by an external resistor tied to the positive supply voltage, and is meant to be switched off during normal operation, when the threshold is properly set and only the binary information at the output of the discriminator is readout.
In all the power consumption measurements presented in the following, the analog buffer was off.

Gain and threshold are programmable thanks to a 16-bit shift register, very similar to a SPI interface.
The first 8 bits control channel 1. Three bits are used to control the gain of the CSA, as will be described in the following, and the remaining five bits control the resistive divider at the noninverting input of the comparator.
The second group of 8 bits controls channel 2.
In this prototype, settings for channels 3 and 4 are copied from those of channels 1 and 2.

The design of the CLARO-CMOS is optimized for negative input charge signals, that is, the ASIC is designed to be used with photodetectors where electrons are collected at the readout electrode.
To accomodate the case where the photodetector signals are made of holes, as for some SiPM models, the same design could be reversed by changing all NMOS transistors with PMOS transistors and viceversa in the CSA, and threshold settings should be changed accordingly.

\subsection{Design of the CSA}

\begin{figure}[t]
\centering
 \def\svgwidth{\linewidth}
 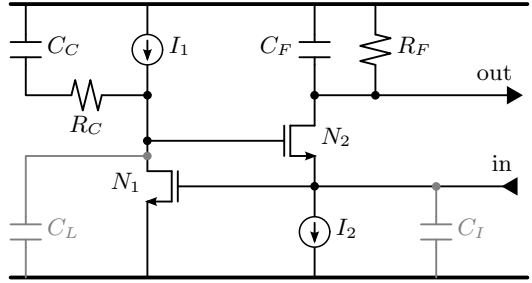
\caption{Simplified schematic of the charge sensitive amplifier (CSA). The parasitic capacitances $C_L$ and $C_I$ are shown in grey. $C_I$ also includes the sensor capacitance.}
\label{schcsasimple}
\end{figure}

Figure \ref{schcsasimple} shows a simplified schematic of the CSA, which includes the parasitic capacitance $C_L$ and the input capacitance $C_I$ for clarity.
The input stage is an active cascode \cite{actcas1, actcas2, actcas3}, a design widely used in the field of photodetector electronics, also referred to as super common base \cite{chase, operaroc, discretesige}.
This design uses a local feedback through $N_1$ to lower the impedance at the source of $N_2$, in order to read the input current pulses on a virtual ground node. The loop gain at intermediate frequency is $g_1 R_C$, where $g_1$ is the transconductance of $N_1$.
The current pulses are integrated by the capacitor $C_F$ at the drain of $N_2$, which discharges through the resistor $R_F$.
The output signal in response to a (negative) charge $Q$ injected at $t=0$ is given by
\begin{equation}
V_O (t) = \frac{Q}{C_F} \frac{\tau_F}{\tau_F - \tau_R} \left( \textrm{e}^{-{t}/{\tau_F}} - \textrm{e}^{-{t}/{\tau_R}}  \right)
\label{signal1}
\end{equation}
where $\tau_R$ is the rise time constant given by the CSA bandwidth and $\tau_F = C_F R_F$ is the fall time constant.
The rise time constant $\tau_R$ is of the order of \mbox{1 ns}, and is directly proportional to $C_I$ as will be shown.
The ASIC is designed for fast photodetectors, where the input current pulse is short, of the order of \mbox{1 ns}. The fall time constant $\tau_F$ was chosen to be \mbox{5 ns}, large enough for an effective integration of fast pulses but small enough to sustain high rates without pile-up.

\begin{figure}[t]
\centering
 \def\svgwidth{\linewidth}
 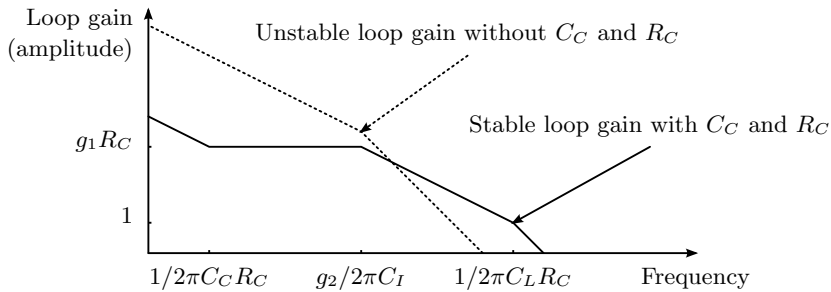
\caption{Bode diagram illustrating the stability of the input feedback loop.}
\label{bodecsa}
\end{figure}

In the simplified scheme of figure \ref{schcsasimple}, the main voltage (or series) noise source is $N_1$ together with the bias circuit $I_1$, while the main current (or parallel) noise source is $R_F$ together with the bias circuit $I_2$.
Transistor $N_2$ contributes to the series noise, but its contribution is divided by the loop gain and becomes negligible. The optimal noise performance corresponds to the case where $N_1$ is biased with a large current $I_1$ to keep its transconductance high and its series noise low. Since $R_F$ contributes to the parallel noise, its value cannot be too small, and this poses an upper limit to the bias current $I_2$ of $N_2$. 
With a low bias current, the transconductance $g_{2}$ of $N_2$ is low, and the input capacitance to ground $C_{I}$ due to the input bonding pad, the bonding wire, packaging, interconnects and to the sensor adds a pole to the input feedback loop at a frequency $g_{2} / 2 \pi C_I$.
If $R_C$ and $C_C$ were not present, the load at the drain of $N_1$ would be purely capacitive, and there would be another pole at very low frequency due to $C_L$. This would be the lower frequency pole of the feedback loop. At the frequency of the second pole, that is $g_{2} / 2 \pi C_I$, the feedback loop would become unstable, unless it were already lower than 1, in which case it would be ineffective in lowering the input impedance at this frequency. This case is illustrated in the bode plot of figure \ref{bodecsa}, dashed line.

To compensate the pole due to $C_L$, $R_C$ and $C_C$ are used. This case is illustrated in the solid line of figure \ref{bodecsa}.
The effect of compensation is to limit the loop gain to $g_1 R_C$ at moderate frequency, higher than $1/2 \pi C_C R_C$. This shifts the pole due to $C_L$ at a higher frequency given by $1/2 \pi C_L R_C$. For this compensation to be effective, it is required that the value of $R_C$ is not too large and that $C_L$ is minimized with a proper layout. In particular, since the area of $C_C$ on silicon is larger than that of $R_C$, its parasitic capacitance to the substrate is larger. A much lower value for $C_L$ is obtained is $R_C$ is placed before $C_C$, as in figure \ref{schcsasimple}.
The relatively low value for $R_C$ strengthens the need to keep high the transconductance of $N_1$, while the transconductance of $N_2$ is less critical.
As illustrated in the solid line of figure \ref{bodecsa}, the dominant pole of the input feedback loop is now at $g_{2} / 2 \pi C_I$. This ensures that the feedback loop is effective in lowering the input inpedance up to a much higher frequency.
The frequency where the loop gain becomes close to unity gives the bandwidth of the CSA. The associated time constant gives the rise time of the output signal:
\begin{equation}
\tau_R  =  \frac{C_I}{g_2} \frac{1}{g_1 R_C +1}   \simeq  \frac{C_I}{g_2 g_1 R_C}
\label{risetime}
\end{equation}
The 10\% to 90\% rise time is given by $2.2\tau_R$.
The rise time is thus directly proportional to the input capacitance $C_I$ and inversely proportional to the loop gain $g_1 R_C$.
The stability of the feedback loop is ensured even if the sensor has a negligible capacitance, since the value of $C_I$ has a lower limit at a few pF due to the gate-drain capacitance of $N_1$, that is less than \mbox{100 fF} but its contribution is multiplied by the loop gain, its gate-source and gate-bulk capacitance (about \mbox{0.5 pF} in total) and the stray capacitance of the pads, the bonding wires, eccetera. Considering all the contributions from the circuit the input capacitance can be estimated to be about $\mbox{1.5 pF}$, bonding pads excluded. With the CLARO-CMOS mounted in a small QFN48 package the total capacitance at the input (without the sensor) was measured to be about \mbox{3.3 pF}.

\begin{figure}[t]
\centering
 \def\svgwidth{\linewidth}
 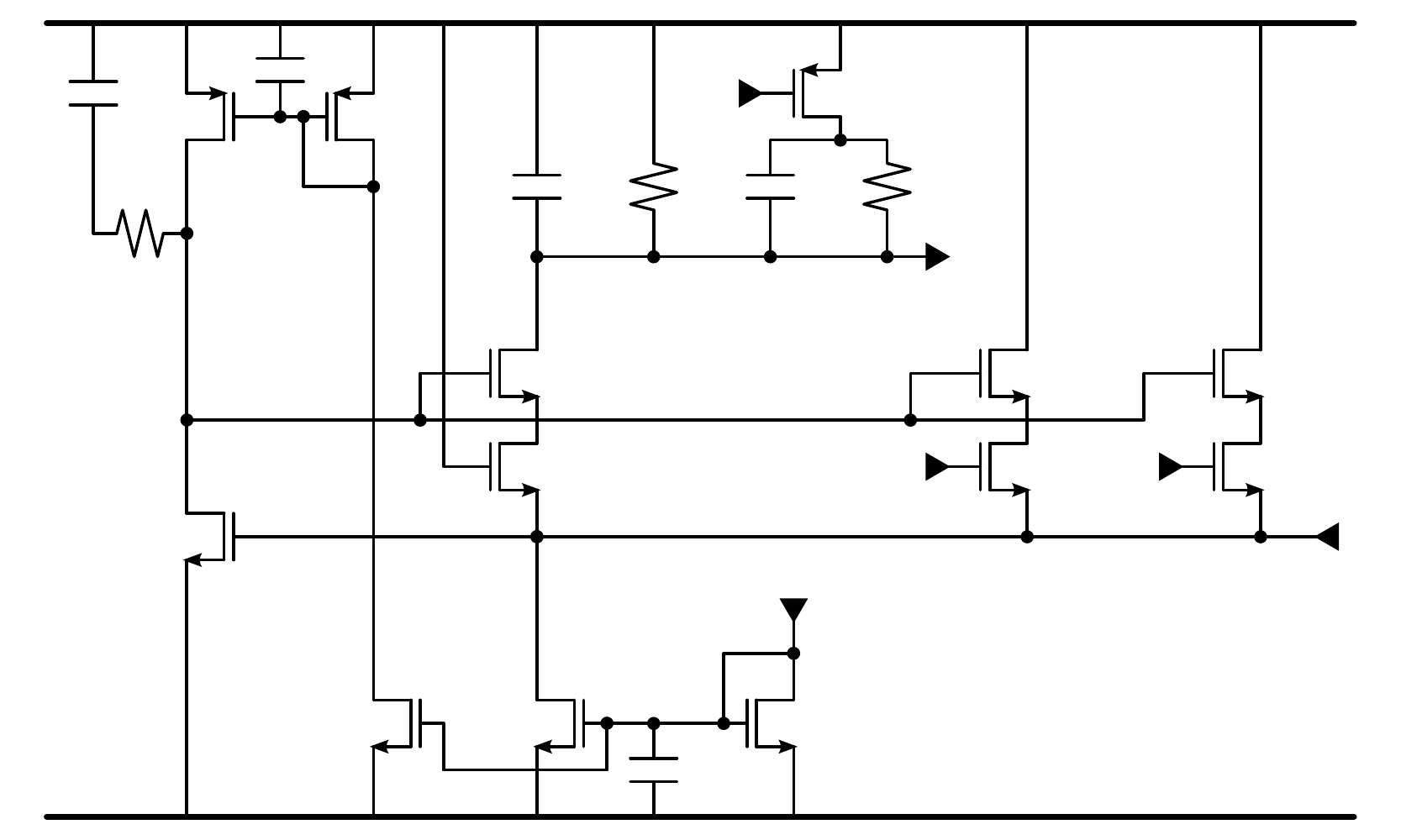
\caption{Full schematic of the CSA. The width of the MOS transistors is shown. The gate length is \mbox{0.35 $\mu$m} for all the transistors in the CSA. The substrate of all NMOS (PMOS) transistors is tied to ground (to the positive rail).}
\label{schcsafull}
\end{figure}

The full schematic of the CSA is shown in figure \ref{schcsafull}.
To vary the gain, a set of MOS switches was included in the design. Two switches, $N_{S3}$ and $N_{S4}$, are used to attenuate the input signal: if the digital control signals $V_3$ or $V_4$ are set high, the switches are closed and a part of the input charge passes through $N_3$ or $N_4$ and is wasted on the positive rail.
The amount of attenuation $B$ is set by choosing the dimensions of $N_3$ and $N_4$, which are 3 and 6 times larger than $N_2$ respectively, causing attenuations of \mbox{$B$ = 4} and \mbox{$B$ = 7}. An attenuation of a factor of \mbox{$B$ = 10} is obtained if both branches are enabled.
The dummy switch $N_{S2}$ whose gate is tied to the positive rail was introduced to preserve the simmetry between the input branches.

Another switch $P_{SF}$ controlled by the digital control signal $V_F$ is used to change the value of $C_F$ and $R_F$, doubling $C_F$ and halving $R_F$, to change the gain by a factor of 2 while keeping the discharge time constant the same.
The voltages $V_3$, $V_4$ and $V_F$ are the three control bits which allow gain setting on each channel.
The reason why only one switch was used to change the values of $C_F$ and $R_F$ is related to the switch parasitics. If several switches were connected in series, their series resistance in the "on" state would have caused distortion in the shape of the output signal. If several of such switches were put in parallel, their capacitance in the "off" state would have been in parallel with $C_F$, reducing the maximum gain achievable.

The dimensions of the bias transistors $N_{B1} \dots N_{B5}$ were chosen so that the bias current of $N_1$ is about 2.5 times larger than that of $N_2$.
Transistor $N_1$ has a very large area to obtain a high transconductance $g_1$.
In this prototype the bias current of the CSA can be set by changing $I_A$ with an external resistor.
Two operating modes were chosen: a ``low power'' mode, with \mbox{$I_A$ = 2 $\mu$A}, and a ``timing'' mode, with \mbox{$I_A$ = 5 $\mu$A}.
In ``low power'' mode, $N_1$ is biased with \mbox{85 $\mu$A}, resulting in \mbox{$g_1$ = 2 mA/V}. Since \mbox{$R_C$ = 10 k$\Omega$}, the low frequency gain of the input feedback loop is about 20. The input branches with $N_2$, $N_3$, $N_4$ are biased with a total current \mbox{25 $\mu$A}. The total transconductance of $N_2$ in parallel with $N_3$ and $N_4$ is about \mbox{350 $\mu$A/V}, depending on which of $N_3$ and $N_4$ are enabled. If the feedback loop were not present, the input impedance would be higher than \mbox{2 k$\Omega$}. The feedback loop lowers this value to about \mbox{130 $\Omega$}.
From equation \ref{risetime} the 10\% to 90\% rise time is expected to be about \mbox{1.2 ns} for \mbox{$C_I$ = 3.3 pF}, and \mbox{2.4 ns} for \mbox{$C_I$ = 6.5 pF}.
In ``timing'' mode, $N_1$ is biased with \mbox{170 $\mu$A}, and its transconductance becomes \mbox{3.8 mA/V}, so that the loop gain roughly doubles. The total transconductance of $N_2$, $N_3$ and $N_4$ is about \mbox{500 $\mu$A/V}. Thanks to the larger loop gain, the input inpedance is now reduced to less than \mbox{100 $\Omega$}. The bandwidth of the CSA is increased, and the loop gain at $1/2 \pi C_L R_L$ becomes closer to unity, but stability of the feedback loop is still ensured even with a negligible sensor capacitance. The rise time of the signal at the output of the CSA as given by equation \ref{risetime} is roughly half than in ``low power'' mode thanks to the larger loop gain. The main consequence is a reduction of the time walk of the discriminator, as will be shown in the following.


The noise of the CSA can be referred to the input as an equivalent noise charge (ENC). The detailed noise calculations are given in appendix \ref{noiseapp}. For \mbox{$\tau_R$ < 0.3 $\tau_F$}, that is for \mbox{$C_I$ < 10 pF} in ``low power'' mode, the ENC is given by
\begin{equation}
\textrm{ENC} \simeq \left(
 i_n^2 \frac{\tau_F + 3 \tau_R }{4} +
e_n^2 C_I^2 \frac{\tau_F + 3 \tau_R }{4 \tau_F \tau_R}  +
A_f  C_I^2 \frac{\tau_F + 4 \tau_R}{\tau_F} \ln{ \frac{\tau_F }{\tau_R}}\right)^{\frac{1}{2}}
\label{enc2}
\end{equation}
where $i_n$ is the current noise density, $e_n$ is the white voltage noise density and $A_f$ is the $1/f$ voltage noise coefficient.
In addition to the noise from $N_1$ and $R_F$, it is necessary to consider the noise contributions coming from the bias transistor $N_{B2}$, whose current noise directly contributes to the parallel noise at the input, and $P_{B5}$, whose current noise is divided by the transconductance of $N_1$ and becomes a series noise contribution at the input.
Moreover, if the value of the filtering capacitors $C_{B2}$ and $C_{B5}$ is not large enough, additional noise coming from $N_{B1}$, $N_{B3}$ and $P_{B4}$ can be injected through $N_{B2}$ and $P_{B5}$, contributing to the parallel and series noise respectively.

In this first CLARO-CMOS prototype, filter capacitors $C_{B2}$ and $C_{B5}$ are not present.
The parallel noise is dominated by the channel current of $N_{B1}$ mirrored and multiplied by 10 by $N_{B2}$. Since in ``low power'' mode the transconductance of $N_{B1}$ is \mbox{$g_{B1}$ = 35 $\mu$A/V} we have
\begin{equation}
i_{B1}^2 = 10^2 \times \frac{8}{3} kT g_{B1}  \simeq \left( 6.2\ \textrm pA/\sqrt{ \textrm Hz} \right)^2
\label{enc3}
\end{equation}
Other contributions come from $N_{B2}$, about \mbox{2 pA/$\sqrt{ \textrm Hz}$}, and from $R_F$, about \mbox{0.9 pA/$\sqrt{ \textrm Hz}$} if \mbox{$V_F$ = 1}, \mbox{1.3 pA/$\sqrt{ \textrm Hz}$} if \mbox{$V_F$ = 0}, assuming $B=1$.
The weight of the noise generated by $R_F$ is directly proportional to the attenuation factor $B$: at the maximum attuenuation, that is with $B=10$, the noise from $R_F$ becomes the dominant parallel noise source with \mbox{9 pA/$\sqrt{ \textrm Hz}$} if \mbox{$V_F$ = 1}, \mbox{13 pA/$\sqrt{ \textrm Hz}$} if \mbox{$V_F$ = 0}. The other noise sources in the CSA do not depend on $B$, since they share the same attenuation as the signal.
Anyway the attenuation is meant to be used only when the signals are large; so in those cases the signal to noise ratio is expected to be anyway adequate.
In the following, for all noise evaluations, we will consider $B=1$.
The sum of all parallel noise is thus close to \mbox{7 pA/$\sqrt{ \textrm Hz}$} in ``low power'' mode with $B=1$.
In ``timing'' mode the parallel noise increases by about 20\% due to the larger bias current which gives a larger transconductance to $N_{B1}$ and $N_{B2}$.

The series noise is dominated by $N_1$ and $P_{B5}$. As already mentioned, additional noise from the other bias transistors is injected through $P_{B5}$ since its gate is not filtered. In ``low power'' mode, where \mbox{$g_1$ = 2 mA/V}, the series white noise is dominated by $N_{B1}$, $N_{B3}$ and $P_{B4}$, which all have a transconductance of about \mbox{$g_{B1}$ = 35 $\mu$A/V}. The resulting white voltage noise at the input is
\begin{equation}
e_{B134}^2 = 25^2 \times 3 \times \frac{8}{3} kT \frac{g_{B1}}{g^2_{1}}  \simeq \left( 13\ \textrm nV/\sqrt{ \textrm Hz} \right)^2
\label{enc4}
\end{equation}
being 25 the area ratio between $P_{B5}$ and $P_{B4}$.
Other contributions come from $N_{1}$, about \mbox{2.3 nV/$\sqrt{ \textrm Hz}$}, and from $P_{B5}$, about \mbox{1.6 nV/$\sqrt{ \textrm Hz}$}. The sum of all series white noise is about \mbox{$e_n$ $\simeq$ 14 nV/$\sqrt{ \textrm Hz}$}.
In ``timing'' mode the series noise reduces by almost a factor of 2, because of the larger transconductance of $N_1$ which gives a larger loop gain.
Compared to the series white noise, the contribution of the $1/f$ component is expected to be negligible since from simulations it is possible to estimate \mbox{$A_f$ < $10^{-9}$ V$^2$}.

According to equations \ref{risetime} and \ref{enc2}, the parallel noise contribution to the ENC at the output of the CSA is expected to be about \mbox{1.8 ke$^-$} (\mbox{0.29 fC}) at \mbox{$C_I$ = 3.3 pF}, and \mbox{2.0 ke$^-$} (\mbox{0.32 fC}) at \mbox{$C_I$ = 6.5 pF}.
The series noise contribution is expected to be about \mbox{7.5 ke$^-$} (\mbox{1.2 fC}) at \mbox{$C_I$ = 3.3 pF}, and \mbox{12 ke$^-$} (\mbox{1.9 fC}) at \mbox{$C_I$ = 6.5 pF}.
The total noise of the CSA is thus expected to be \mbox{7.7 ke$^-$} (\mbox{1.2 fC}) at \mbox{$C_I$ = 3.3 pF}, and \mbox{12 ke$^-$} (\mbox{1.9 fC}) at \mbox{$C_I$ = 6.5 pF}.
At the auxiliary output, the rise time is limited by the bandwidth of the analog buffer. In that case the weight of the series noise is expected to be smaller, while the weight of the parallel noise is expected to be larger, according to equation \ref{enc2}.
For instance, assuming that the output buffer limits the output signals with time constants of \mbox{$\tau_R$ = 1.3 ns} and \mbox{$\tau_F$ = 7.2 ns}, equation \ref{enc2} gives \mbox{5.6 ke$^-$} (\mbox{0.89 fC}) with an input capacitance of \mbox{3.3 pF}, dominated by the series noise.

As already discussed, the filtering capacitors $C_{B2}$ and $C_{B5}$ can be used to improve the noise performance of the design, considerably reducing both the series and the parallel noise injected through the bias transistors, at the price of a larger layout area on silicon. This improvement will be considered for the next versions of the ASIC.

\subsection{Design of the comparator}

\begin{figure}[t]
\centering
 \def\svgwidth{\linewidth}
 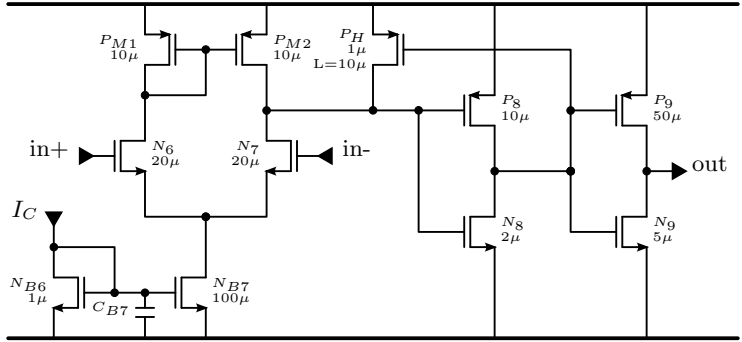
\caption{Full schematic of the comparator. The width of the MOS transistors is shown. The gate length is \mbox{0.35 $\mu$m} for all the transistors except $P_{H}$. The substrate of all NMOS (PMOS) transistors is tied to ground (to the positive rail).}
\label{schcompfull}
\end{figure}

Figure \ref{schcompfull} shows the schematic of the comparator.
The input stage is a differential pair loaded with a current mirror. This is the only part of the comparator which dissipates a continuous current.
Since $I_C$ is about \mbox{1 $\mu$A}, the differential pair is biased with about \mbox{100 $\mu$A}.
The signal from the CSA is connected to the inverting input of the comparator, while the noninverting input is held at a constant potential which defines the threshold.
The threshold voltage at the inverting input of the discriminator can be set between \mbox{1.25 V} (half the positive rail voltage) and \mbox{0.83 V} (one third the positive rail voltage) in 32 steps, labelled from 0 to 31, thanks to a 5-bit DAC implemented as a simple voltage divider.
Each step is about \mbox{13 mV}. At the maximum gain, this corresponds to a threshold step of \mbox{150 ke$^-$} (\mbox{24 fC}).

In ready state, the output of the differential pair is low, and stays close to \mbox{0.5 V}. This signal feeds the inverter made of $P_8$ and $N_8$. Transistor $N_8$ is small and has a large threshold, about \mbox{0.6 V}. In this way $N_8$ is biased just below threshold: no current passes through the first inverter and its output is high.
Transistor $P_H$ provides hysteresis, and since its gate is high it is switched off.
The output of $P_8$ and $N_8$ is fed to the second inverter made of $P_9$ and $N_9$, which is also the output stage.

In response to a negative pulse from the CSA, the output of the differential pair goes up, close to the positive rail. The output of first inverter goes to ground, closing the switch $P_H$, which draws current from the differential pair and holds up its output providing hysteresis. At the same time, the output of $P_9$ and $N_9$ swings to the positive rail. 
The gate length of $P_H$ is large: its ``on'' resistance is about 150 k$\Omega$, so that only a fraction of the bias current of the differential pair passes through $P_H$, and after a few nanoseconds the output of the differential pair is able to get back to the initial condition. 
When the output of the differential pair goes down, the output of the first inverter goes up, transistor $P_H$ is opened and the output of the comparator goes down.
After this the discriminator is ready to trigger another pulse from the CSA. The width of the output pulses is proportional to the amplitude of the input signals, allowing to apply time over threshold algorythms to determine the input charge and compensate for time walk.

The gain of the input stage of the comparator is about \mbox{30 V/V} for small signals around threshold at low frequency, with a pole at about \mbox{30 MHz}. The corresponding time constant is \mbox{$\tau_C$ $\simeq$ 5 ns}, about the same as the fall time of the CSA pulse $\tau_F$. The effect of hysteresis is to increase the gain to \mbox{600 V/V} at low frequency. The gain of the inverters is about \mbox{20 V/V} for each. The overall gain of the comparator at low frequency including hysteresis results in \mbox{24 $\times$ $10^4$ V/V} or \mbox{107 dB}.
Transistor $P_9$ is much larger than $N_9$, in order to obtain a very fast transition on the rising edge at the output.
The rise and fall times of the output signal depend on the load at the output of the discriminator. The output stage was designed to drive only a short line to a digital processing circuit or to an external low impedance driver, located a few cm away on the same board. Thus a purely capacitive load of a few pF is expected. This was done in order to give the maximum flexibility in the design of a full system and to avoid unnecessary power consumption in the CLARO-CMOS.
The output signal is limited by the slew rate of the output stage on the output load, that is $I_L / C_L$, where $I_L$ is the current from the output stage, and $C_L$ is the output load capacitance. The output current can be estimated to be $I_L \simeq$ 2.5 V$ \times g_9$, where $g_9$ is the transconductance of the output stage.
For small signals, the transconductance of $P_9$ is \mbox{2 mA/V}, and that of $N_9$ is \mbox{0.8 mA/V}, even if this values are largely non linear since the output stage swings from rail to rail. Anyway the rise time is expected to be about two times smaller than the fall time, since the rising edge is driven by $P_9$ while the falling edge is driven by $N_9$. With these numbers, the time required for the full swing from \mbox{0 V} to \mbox{2.5 V} at the output is about
${2.5\ \textrm{V}}/({I_L / C_L}) \simeq {C_L}/{g_9}$.
With a load capacitance of \mbox{$C_L$ = 8 pF}, for instance, the output 0\% to 100\% rise time is \mbox{4 ns}, which corresponds to a 10\% to 90\% rise time of \mbox{3.2 ns}, and the output 100\% to 0\% fall time is \mbox{10 ns}, which corresponds to a 90\% to 10\% fall time of \mbox{8 ns}.

The input transistors $N_6$ and $N_7$ have a transconductance $g_C$ of about \mbox{700 $\mu$A/V}, while $P_{M1}$ and $P_{M2}$ have a transconductance $g_M$ of about \mbox{300 $\mu$A/V}. These are the main contributors to the noise of the comparator. Transistor $N_{B7}$ does not contribute because its noise is common mode while the input stage is differential. So in the case of the comparator the bias filtering capacitor $C_{B7}$ can be avoided. The input referred white voltage noise density can be expected to be
\begin{equation}
e_{C}^2 = 2 \times \frac{8}{3} kT \frac{1}{g_{C}} + 2 \times \frac{8}{3} kT \frac{g_{M}}{g^2_{C}}  \simeq \left( 6.7\ \textrm nV/\sqrt{ \textrm Hz} \right)^2
\end{equation}
which together with the $1/f$ contributions corresponds to a voltage noise at the input of about \mbox{65 $\mu$V} RMS.
Compared with the RMS noise at the output of the CSA, that is more than \mbox{1 mV} RMS in the best case of a \mbox{3.3 pF} input capacitance, this contribution is negligible, at least with the attenuation factor $B=1$. With larger attenuations the weight of the noise of the comparator grows accordingly, and at $B=10$ it becomes significant. Since as already mentioned the attenuation is only meant to be used with very large signals, where the signal to noise ratio is a minor concern, we will anyway consider the case of $B=1$ in the following.
The jitter on the rising edge of the comparator is expressed by
\begin{equation}
\sigma_{\textrm{\footnotesize \ Rise}} \simeq \frac{\tau_C}{ Q-Q_{TH} }
\left(i_n^2  \frac{\tau_C}{8} + 
e_n^2 C_I^2 \frac{1}{8 \tau_C} +
A_f C_I^2   \frac{1}{2}
 \right)^{\frac{1}{2}}
\label{jittermain1}
\end{equation}
The calculations to obtain equation \ref{jittermain1} are reported in appendix \ref{jitterapp}. The time constant \mbox{$\tau_C$ $\simeq$ 5 ns} is given by the bandwidth of the first stage of the comparator. When the threshold is set at \mbox{300 ke$^-$} (\mbox{48 fC}), equation \ref{jittermain1} predicts a jitter of \mbox{32 ps} for \mbox{600 ke$^-$} (\mbox{96 fC}) signals, of which \mbox{24 ps} are due to the series noise, and \mbox{18 ps} to the parallel noise. As for the case of the ENC, the $1/f$ component is negligible. According to equation \ref{jittermain1}, jitter is expected to decrease to \mbox{8 ps} for \mbox{1.5 Me$^-$} (\mbox{240 fC}) signals.
For larger signals, equation \ref{jittermain1} predicts an unlimited improvement; in reality the slope of the signal at the first stage of the discriminator is also limited by slew rate. So, in contrast with equation \ref{jittermain1}, jitter is expected at some point to stop decreasing for larger signals, and to saturate to a constant value.

\section{Performance of the prototype}

\begin{figure}[t]
\centering
\def\svgwidth{\linewidth}
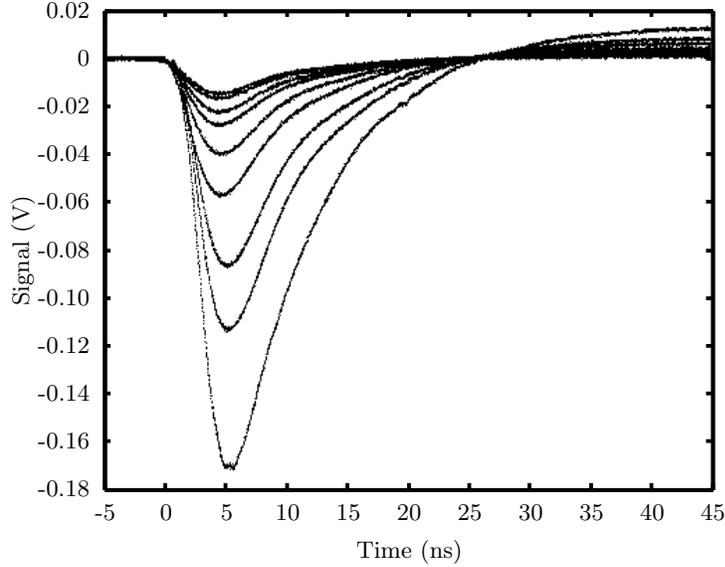
\caption{Signals at the output of the analog buffer (auxiliary output) in ``low power'' mode.}
\label{csasignal}
\end{figure}

Figure \ref{csasignal} shows the signal at the output of the CSA in ``low power'' mode, read out at the auxiliary output through the PMOS follower biased with a \mbox{1 k$\Omega$} resistor to the positive rail.
The gain was set to the maximum value (\mbox{$V_3$ = $V_4$ = 0}, \mbox{$V_F$ = 1}), and pulses from \mbox{330 ke$^-$} (\mbox{53 fC}) to \mbox{3.3 Me$^-$} (\mbox{530 fC}) were injected at the input by a Agilent 81130A \mbox{600 MHz} step generator through a \mbox{0.5 pF} test capacitance.
The 10\% to 90\% rise time of the test signals is \mbox{0.6 ns}, simulating the typical charge collection time of a fast photomultiplier.
The output of the PMOS follower was buffered with a Texas Instruments LMH6703 fast opamp driving a terminated \mbox{50 $\Omega$} line.
The signals were acquired with a Agilent DCA-X 86100D \mbox{20 GHz} sampling scope with the bandwidth limited to \mbox{12 GHz} in our measurements.

The leading edge of the measured analog signal in response to a \mbox{330 ke$^-$} (\mbox{53 fC}) pulse is \mbox{2.8 ns} (10\% to 90\%), its trailing edge is {15.8 ns} (90\% to 10\%), the pulse width at 50\% is {8 ns}.
The corresponding time constants are \mbox{$\tau_R$ = 1.3 ns} and \mbox{$\tau_F$ = 7.2 ns}.
Due to the finite bandwidth of the PMOS follower, the measured signal is slower than the signal at the output of the CSA which feeds the input of the discriminator.
Since the transconductance of the PMOS follower is less than \mbox{1 mA/V} and its bias resistor is \mbox{1 k$\Omega$}, the amplitude of the buffered signal is smaller than at the output of the CSA.

The input noise was obtained by measuring the baseline noise at the auxiliary output and referring it to the input of the CSA as an equivalent noise charge (ENC). The measured ENC for an input capacitance of \mbox{3.3 pF} is \mbox{6 ke$^-$} (\mbox{1 fC}) RMS, consistent with equation \ref{enc2}, which predicts \mbox{5.6 ke$^-$} (\mbox{0.89 fC}), as already mentioned, once the correct rise and fall time measured at the output of the analog buffer are considered.
The importance of low noise is mainly related with timing performance, which will be discussed in the following.

\begin{figure}[t]
\centering
\def\svgwidth{\linewidth}
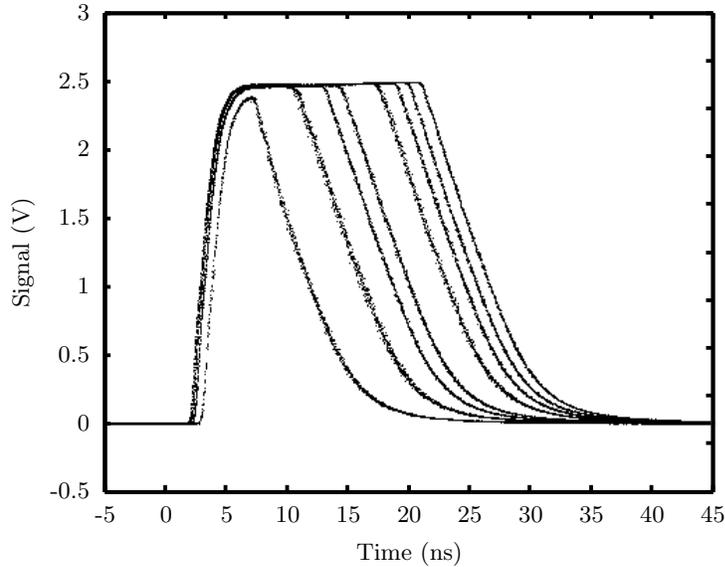
\caption{Signals at the output of the discriminator (main output) in ``low power'' mode.}
\label{compsignal}
\end{figure}

Figure \ref{compsignal} shows the signal at the output of the discriminator when the CLARO-CMOS is operated in ``low power'' mode. The threshold was set at 6, which at the maximum gain corresponds to 800 ke$^-$ (\mbox{128 fC}), and signals from \mbox{810 ke$^-$} (\mbox{130 fC}) to \mbox{5.6 Me$^-$} (\mbox{900 fC}) were injected at the input. This range of input signals corresponds to the typical single photon response of a photomultiplier in nominal bias condition.
As altready mentioned, the output stage of the discriminator is designed to drive a capacitive load of a few pF. In these tests the capacitive load at the output was measured to be \mbox{8 pF}, contributed by the pads, the QFN48 package, and a short (a few cm) PCB trace to a Texas Instruments LMH6703 fast opamp used as a low impedance driver to the sampling scope. With this load, the 10\% to 90\% rise time is \mbox{2.2 ns}, and the 90\% to 10\% fall time is \mbox{9.3 ns}.
The 50\% pulse width depends on the amount of charge injected at the input, ranging from \mbox{7.2 ns} for the shortest signal in figure \ref{compsignal}, that is just above threshold, to \mbox{21.7 ns} for the largest signal in figure \ref{compsignal}, that is almost a factor of 10 above threshold.
The delay between the input charge pulse and the time when the output of the discriminator reaches 50\% is \mbox{5 ns} for signals just above threshold, and lowers to about \mbox{2.5 ns} for signals well above threshold. The delay is due to the rise time of the CSA pulse at the input of the comparator and to the difference in the speed of the comparator for different levels of overdrive. The difference between the two extreme values, about \mbox{2.5 ns} in ``low power'' mode, constitutes the time walk of the discriminator, which is critical for timing performance, to be discussed in the following.

This performance was obtained in ``low power'' mode, with an overall continuous power dissipation per channel of \mbox{0.7 mW}.
If the discriminator is triggered with a \mbox{10 MHz} rate, the average power consumption increases to \mbox{1.9 mW} per channel.
It is worth noting that the signals in figures \ref{csasignal} and \ref{compsignal} are acquired at the output of the sampling scope: the displayed signals are obtained as the superposition of dots from several output signals, while the sampling trigger was synchronized with the step generator.
In this way the figure incorporates at a glance also noise and jitter.
The output signals shown demonstrate the capability of the CLARO-CMOS to count fast pulses from photomultipliers, from the single photoelectron up to larger gains, with a low noise, very high rate (up to \mbox{10 MHz}), and a very low power consumption.

When the prototype is operated in ``timing'' mode, the power consumption is increased to \mbox{1.5 mW} per channel (rising to \mbox{2.3 mW} per channel with a \mbox{10 MHz} rate). The difference in the output signals between ``low power'' and ``timing'' modes are small: the different power consumption affects only the output of the CSA, but the difference cannot be directly appreciated on the shape of the buffered signals because of the bandwidth limitation of the auxiliary output buffer. The differences between the two operating modes can be appreciated on the crosstalk and jitter measurements presented in the following.

\subsection{Crosstalk}

\begin{figure}[t]
\centering
\def\svgwidth{\linewidth}
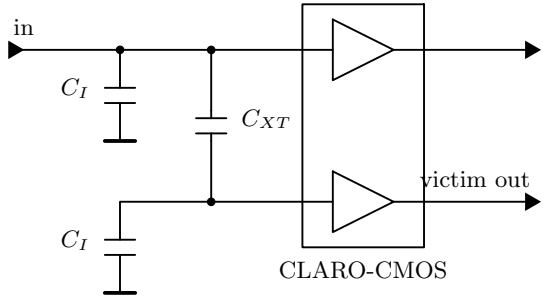
\caption{Setup for crosstalk measurement. The capacitance $C_{XT}$ represents the stray capacitance between the pixels of the sensor.}
\label{crosstalksetup}
\end{figure}

\begin{figure}[t]
\centering
\def\svgwidth{\linewidth}
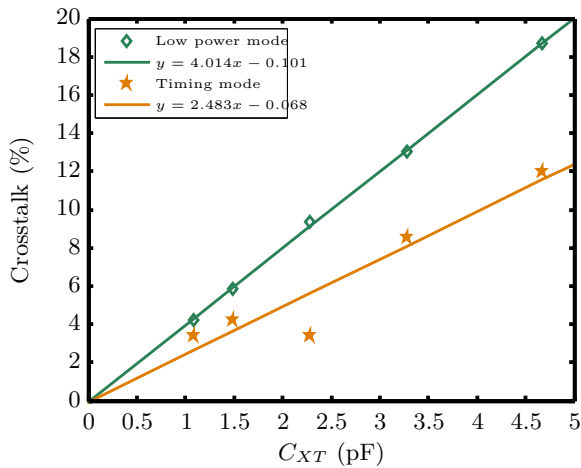
\caption{Crosstalk versus crosstalk capacitance $C_{XT}$ in ``low power'' and ``timing'' modes.}
\label{crosstalkvscxt}
\end{figure}

With very fast circuits, such as the CLARO-CMOS, crosstalk may be critical. Fast signals could be capacitively coupled to neighbouring channels through parasitic capacitances much more easily than with slower circuits.
The level of crosstalk between channels was measured as follows. The gain of the victim channel was set to the maximum value and its threshold was set at \mbox{300 ke$^-$} (\mbox{48 fC}). No signal was applied at the input of the victim, while large signals were injected at the input of a neighbouring channel. The crosstalk could be estimated from the amplitude of the minimum signal which triggers the discriminator of the victim.
To simulate the real case where different pixels of a pixellated photodetector are connected to the inputs of the CLARO, a crosstalk capacitance $C_{XT}$ was added between the inputs as depicted in figure \ref{crosstalksetup}. The input capacitance to ground in this measurement was \mbox{$C_I$ = 6.5 pF}.

The level of crosstalk was measured with different values of $C_{XT}$ both in ``low power'' and ``timing'' modes, and the results are plotted and linearly fitted in figure \ref{crosstalkvscxt}.
The crosstalk found on chip, that is with \mbox{$C_{XT}$ = 0}, is negligible. Signals up to \mbox{10 Me$^-$} (\mbox{1.6 pC}) where injected without triggering the victim.
Increasing the value of \mbox{$C_{XT}$} causes the crosstalk to increase correspondingly. The measured data were fitted with lines, whose intercept value is compatible with zero, confirming that no crosstalk is observed if no capacitance is added outside the ASIC between the inputs.
The value of \mbox{$C_{XT}$} in a given application depends on the type of sensor. For instance, the capacitance between the anodes of a Hamamatsu R7600 Ma-PMT is less than \mbox{0.5 pF}. This would translate in a crosstalk level below 2\% in ``low power'' mode, and below 1\% in ``timing'' mode.
A lower level of crosstalk is obtained in ``timing'' mode thanks to the lower input impedance, due to the larger loop gain in the CSA as already discussed.
For fast readout of pixellated sensors it is mandatory that the parasitic capacitance between neighbouring inputs is kept under control. In the cases where the capacitance $C_{XT}$ cannot be reduced due to the characteristics of the sensor, a larger $C_{I}$ should be used. This would affect noise and bandwidth, but would help in eliminating crosstalk.

\subsection{Timing resolution}

To evaluate the timing performance of the CLARO-CMOS prototype the gain of the CSA was set to the maximum value, and the threshold of the discriminator was set at \mbox{300 ke$^-$} (\mbox{48 fC}).
Since the timing performance is expected to be directly proportional to the signal to noise ratio, the use of small input signals corresponds to a conservative, worst case scenario.
The time resolution of this setup was estimated to be \mbox{7 ps} RMS by directly connecting the Agilent 81130A step generator to the Agilent DCA-X 86100D sampling scope. Some of the measurements presented in the following reach \mbox{10 ps}: in these cases the result is partially limited by the setup.
The setup contribution of \mbox{7 ps} was subtracted in quadrature from the measurements.
Moreover, as already mentioned, the 10\% to 90\% rise time of the input test signals is \mbox{0.6 ns}, which is not negligible compared to the rise time predicted at the output of the CSA by equation \ref{risetime} in ``timing'' mode and with a low input capacitance.
As expressed by equation \ref{jittermain1}, the timing resolution on the rising edge of the discriminator signal is limited by the time contant of the first stage of the comparator $\tau_C$ about \mbox{5 ns}. Thus the contribution of \mbox{0.6 ns} due to the test signal generator is expected to be negligible in the jitter measurements.
It may anyway have some impact on the effectiveness in time over threshold compensation presented in the following.

\begin{figure}[t]
\centering
\def\svgwidth{\linewidth}
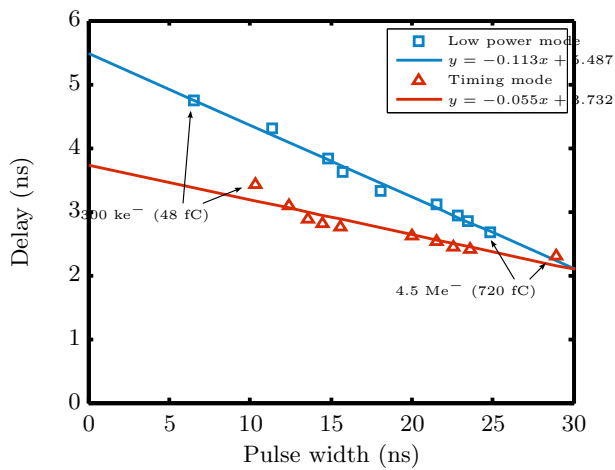
\caption{Delay versus pulse width in ``low power'' and ``timing'' modes.}
\label{delayvspulsewidth}
\end{figure}

The overall timing performance of a system composed of a sensor and a low jitter readout circuit depends also on the precision of time walk compensation; otherwise the low jitter would be spoiled by the time walk induced by the amplitude spread of the signals coming from the sensor.
Figure \ref{delayvspulsewidth} shows the dependence of the delay on the pulse width, starting from signals just above threshold.
The difference in the delay for a given range of input charge is the time walk of the discriminator.
This is the fundamental curve on which the time walk compensation based on time over threshold measurement is based. 
The slope of the fitting lines can be used to estimate the time over threshold effectiveness in compensating time walk.
To a first order approximation, the curves of figure \ref{delayvspulsewidth} do not depend on threshold.
The measurements were taken both in ``low power'' mode and ``timing'' mode. In ``low power'' mode, as already mentioned, the delay ranges from about \mbox{5 ns} to \mbox{2.5 ns}, thus the time walk for this range of input signals, that is the difference between the two, is \mbox{2.5 ns}. 
In ``timing'' mode, as shown in figure \ref{delayvspulsewidth}, the time walk of the discriminator reduces by about a factor of 2. Thus, even if the shape of the output signals and the maximum sustainable rate are the same as in ``low power'' mode, the effectiveness of a time over threshold measurement in compensating time walk is improved by a factor of 2.

\begin{figure}[t]
\centering
\def\svgwidth{\linewidth}
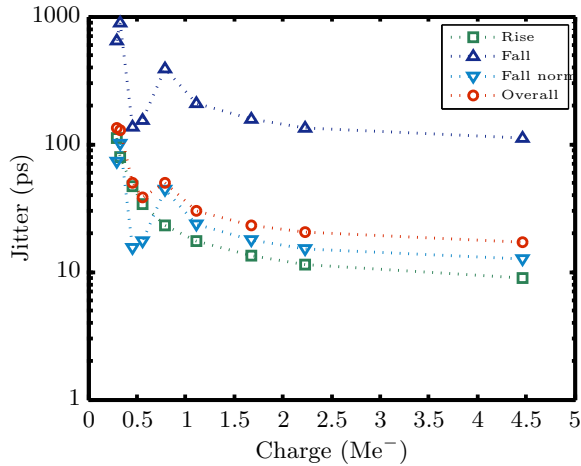
\caption{Jitter versus input charge in ``low power'' mode.}
\label{jittervschargelp}
\end{figure}

The measured RMS jitter versus input charge is displayed in figure \ref{jittervschargelp} for the ``low power'' mode. The plot shows the jitter on the rising edge, that is \mbox{113 ps} on threshold (about \mbox{300 ke$^-$}, or \mbox{48 fC}), decreasing to \mbox{34 ps} for signals of \mbox{560 ke$^-$} or \mbox{90 fC} and then reaching \mbox{9 ps} for large signals (\mbox{4.5 Me$^-$}, or \mbox{720 fC}).
The measured values are in a good match with the values predicted by equation \ref{jittermain1}.
For larger pulses, the rising edge jitter stops decreasing and saturates to a constant value.

The jitter on the falling edge is larger because the transition is slower. Moreover, the jitter on the falling edge is affected by a small disturbance which occurs on ground when the discriminator triggers. This explains the non-monotonic behaviour of the falling edge jitter shown in figure \ref{jittervschargelp}. Anyway, the falling edge is only used to compensate time walk: thus the weight of the falling edge jitter on a timing measurement is given by relation between time walk and pulse width, that is the slope $\gamma$ of the lines used to fit the data in figure \ref{delayvspulsewidth}. In other words, the jitter on the falling edge is normalized according to
\begin{equation}
\sigma_{\textrm{\footnotesize \ Fall\ norm}} = \gamma \ \sigma_{\textrm{\footnotesize \ Fall}}
\end{equation}
where $\gamma$ is 0.113 in ``low power'' mode and 0.055 in ``timing'' mode, as shown in the legend of figure \ref{delayvspulsewidth}.
The jitter on the falling edge normalized with this weight is shown in the plot, and is about 100 ps just above threshold, decreasing to 13 ps with large signals. The overall timing performance (including time walk compensation) is given by the quadratic sum of the rising edge jitter and the normalized falling edge jitter, and is shown in the red curve of figure \ref{jittervschargelp}, going from \mbox{135 ps} just above threshold to \mbox{50 ps} at \mbox{780 ke$^-$} (\mbox{125 fC}), furtherly decreasing to \mbox{17 ps} with \mbox{4.5 Me$^-$} (\mbox{720 fC}) signals.

\begin{figure}[t]
\centering
\def\svgwidth{\linewidth}
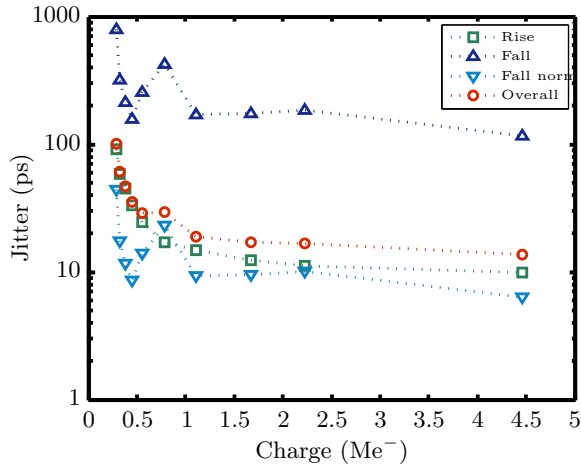
\caption{Jitter versus input charge in ``timing'' mode.}
\label{jittervschargetm}
\end{figure}

The same measurements are given in figure \ref{jittervschargetm} for the ``timing'' mode. The RMS jitter on the rising edge goes from \mbox{92 ps} just above threshold (\mbox{300 ke$^-$}, or \mbox{48 fC}) to \mbox{10 ps} with large signals (\mbox{4.5 Me$^-$}, or \mbox{720 fC}).
Now the rise time $\tau_R$ of the CSA pulse is smaller than in ``low power'' mode, so the jitter on the rising edge is a bit smaller than in ``low power'' mode, but since the speed is in any case limited by the first stage of the discriminator the values are still in agreement with the values predicted by equation \ref{jittermain1}.
Since now the time walk compensation is twice as effective than before, the normalized jitter on the falling edge goes from \mbox{44 ps} to \mbox{6 ps}, becoming almost negligible. The overall timing resolution is thus \mbox{102 ps} just above threshold, quickly decreasing below \mbox{50 ps} above \mbox{380 ke$^-$} (\mbox{61 fC}), and ultimately reaching \mbox{14 ps} for \mbox{4.5 Me$^-$} (\mbox{720 fC}) signals.

\section{Conclusions}

The first protototype of the CLARO-CMOS was deeply characterized with a particular emphasys on its timing resolution, also considering the effectiveness of time walk compensation through time over threshold measurement.
The prototype performes as expected, proving the adequacy of the design approach described.
The obtained time resolution down to \mbox{10 ps} RMS for input charge pulses corresponding to single photoelectron signals from a typical photomultiplier is outstanding, considering the very low power dissipation of the prototype, below \mbox{1 mW} per channel.

\section*{Acknowledgements}

The authors warmly thank Angelo Cotta Ramusino and Roberto Malaguti from INFN Sezione di Ferrara for taking care of the layout and realization of the CLARO-CMOS test board, with which all the characterization and tests presented in this paper were performed.

\appendix

\section{Calculations}

\subsection{Noise}
\label{noiseapp}

In the complex frequency domain the transfer function of the CSA of figure \ref{schcsasimple} can be written as
\begin{equation}
\textrm{TF}_{\textrm{\footnotesize \ CSA}}  (s) = \frac{1}{s C_F} \frac{s \tau_F}{\left( 1 + s \tau_F \right) \left( 1 + s \tau_R \right)}
\label{signals1}
\end{equation}
where $s = i \omega$ is the complex frequency, $\tau_F = C_F R_F$, and $\tau_R = C_I / g_2 g_1 R_C$ as given by equation \ref{risetime}.
The response to a delta-like pulse $Q \delta (t)$ is obtained by multiplying equation \ref{signals1} by $Q$ and taking the inverse laplace transform, which gives equation \ref{signal1}.
A white current noise density $i_n$ at the input is converted to a voltage noise at the output which is given by
\begin{equation}
V_{O i} (s) = \frac{i_n}{s C_F} \frac{s \tau_F}{\left( 1 + s \tau_F \right) \left( 1 + s \tau_R \right)}
\label{rmsout1}
\end{equation}
A voltage white noise density $e_n$ at the input can be converted to its Norton equavalent, that is a current noise density $s C_I e_n$.
The corresponding voltage noise density at the output is
\begin{equation}
V_{O e} (s) = e_n \frac{C_I}{C_F} \frac{s \tau_F}{\left( 1 + s \tau_F \right) \left( 1 + s \tau_R \right)}
\label{rmsout2}
\end{equation}
and the same happens for a voltage low frequency noise $A_f / f$, which gives
\begin{equation}
V_{O A} (s) = \frac{A_f}{f} \frac{C_I}{C_F} \frac{s \tau_F}{\left( 1+ s \tau_F \right) \left( 1 + s \tau_R \right)}
\label{rmsout3}
\end{equation}
To obtain the squared RMS noise at the output, one must integrate the squared amplitudes of equations \ref{rmsout1}, \ref{rmsout2} and \ref{rmsout3} in $d \omega / 2 \pi$ over the whole frequency spectrum.
Equation \ref{rmsout1} gives
\begin{equation}
V_{O i\ \textrm{\footnotesize RMS}}^2 =  \frac{i_n^2}{C_F^2} 
\int_0^{\infty} \frac{\tau_F^2}{\left( 1 + \omega^2 \tau_F^2 \right) \left( 1 + \omega^2 \tau_R^2 \right)}   \frac{d \omega}{2 \pi}
=
\frac{i_n^2}{C_F^2} \frac{\tau_F^2}{4 \left(  \tau_F + \tau_R \right)}
\label{rmsout4}
\end{equation}
equation \ref{rmsout2} gives
\begin{equation}
V_{O e\ \textrm{\footnotesize RMS}}^2 =  e_n^2 \frac{C_I^2}{C_F^2} 
\int_0^{\infty} \frac{\omega^2 \tau_F^2}{\left( 1 + \omega^2 \tau_F^2 \right) \left( 1 + \omega^2 \tau_R^2 \right)}   \frac{d \omega}{2 \pi}
=
e_n^2 \frac{C_I^2}{C_F^2}  \frac{\tau_F^2}{4 \left( \tau_F^2 \tau_R + \tau_F \tau_R^2 \right)}
\label{rmsout5}
\end{equation}
and equation \ref{rmsout3} gives
\begin{equation}
V_{OA\ \textrm{\footnotesize RMS}}^2 =  A_f \frac{C_I^2}{C_F^2}
\int_0^{\infty} \frac{\omega \tau_F^2}{\left( 1 + \omega^2 \tau_F^2 \right) \left( 1 + \omega^2 \tau_R^2 \right)}  d \omega
=
A_f \frac{C_I^2}{C_F^2} \frac{\tau_F^2   }{ \tau_F^2   - \tau_R^2 }  \ln{ \frac{\tau_F }{\tau_R}}
\label{rmsout6}
\end{equation}
If one lets $\tau_R \rightarrow \tau_F$ the above expressions reduce to the known expression for a RC-CR filter \cite{gattimanfredi}.
In our case $\tau_R \leq 0.3 \tau_F $, so we can approximate expanding to the first order in $\tau_R / \tau_F$.
Equation \ref{rmsout4} becomes
\begin{equation}
V_{O i\ \textrm{\footnotesize RMS}}^2 \simeq 
\frac{i_n^2}{C_F^2} \frac{\tau_F - \tau_R }{4}
\label{rmsout7}
\end{equation}
equation \ref{rmsout5} becomes
\begin{equation}
V_{O e\ \textrm{\footnotesize RMS}}^2 \simeq 
e_n ^2 \frac{C_I^2}{C_F^2} \frac{\tau_F - \tau_R }{4 \tau_F \tau_R}
\label{rmsout8}
\end{equation}
and equation \ref{rmsout6} becomes
\begin{equation}
V_{O A\ \textrm{\footnotesize RMS}}^2 \simeq 
A_f  \frac{C_I^2}{C_F^2}  \ln{ \frac{\tau_F }{\tau_R}}
\label{rmsout9}
\end{equation}
Summing together equations \ref{rmsout7}, \ref{rmsout8} and \ref{rmsout9} one obtains the total squared RMS noise at the output:
\begin{equation}
V_{O \ \textrm{\footnotesize RMS}}^2 \simeq \frac{i_n^2}{C_F^2} \frac{\tau_F - \tau_R }{4} + e_n^2 \frac{C_I^2}{C_F^2} \frac{\tau_F - \tau_R }{4 \tau_F \tau_R} + A_f  \frac{C_I^2}{C_F^2}  \ln{ \frac{\tau_F }{\tau_R}}
\label{rmsout10}
\end{equation}
The square root of equation \ref{rmsout10} gives the total RMS noise at the output of the CSA.
To obtain the noise referred to the input as ENC one must calculate
\begin{equation}
\textrm{ENC} = \frac{Q}{V_{O\ \textrm{\footnotesize MAX}}(Q)} V_{O \ \textrm{\footnotesize RMS}}
\label{encapp1}
\end{equation}
where $V_{O\ \textrm{\footnotesize MAX}}(Q)$ is the peak output voltage for a charge $Q$, which can obtained from equation \ref{signal1} and is
\begin{equation}
V_{O\ \textrm{\footnotesize MAX}}(Q) = \frac{Q}{C_F} \frac{\tau_F}{\tau_F - \tau_R}  \left[
\left( \frac{\tau_R}{\tau_F}  \right)^{\frac{\tau_R}{\tau_F - \tau_R}} -
\left( \frac{\tau_R}{\tau_F}  \right)^{\frac{\tau_F}{\tau_F - \tau_R}}
\right]
\label{vomax1}
\end{equation}
which expanding for small $\tau_R / \tau_F$ becomes
\begin{equation}
V_{O\ \textrm{\footnotesize MAX}}(Q) \simeq \frac{Q}{C_F} \left[ 1 + \frac{\tau_R}{\tau_F}  + \left( \frac{\tau_R}{\tau_F} \right)^2  \right]
\left[
\left( \frac{\tau_R}{\tau_F}  \right)^{\frac{\tau_R}{\tau_F}}
\left( 1 - \frac{\tau_R}{\tau_F} \right)
\right]
 \simeq 
\frac{Q}{C_F}
\left(
1 + \frac{\tau_R}{\tau_F} \ln{\frac{\tau_R}{\tau_F}}
\right)
\label{vomax2}
\end{equation}
where the expression was approximated using the fact that $x^x \simeq 1 + x \ln x$ for small $x$, and all the terms in $\tau_R / \tau_F$ with power equal or higher than 2 were dropped.
Equation \ref{vomax2} can be furtherly approximated by
\begin{equation}
V_{O\ \textrm{\footnotesize MAX}}(Q) \simeq \frac{Q}{C_F} \left(1+2 \frac{\tau_R}{\tau_F} \right)^{-1}
\label{vomax3}
\end{equation}
For $\tau_R \ll \tau_F$, both equations give $V_{O\ \textrm{\footnotesize MAX}}(Q) = {Q}/{C_F}$. For $\tau_R \leq 0.3 \tau_F$, equation \ref{vomax3} approximates equation \ref{vomax2} within a 10\% error. For instance, for $\tau_R = \tau_F / \textrm{e}$ equation \ref{vomax2} gives $0.63\ {Q}/{C_F}$, while equation \ref{vomax3} gives $0.58\ {Q}/{C_F}$. The approximation of equation \ref{vomax3} is based on the fact that the coefficient 2 is the closest integer to $\textrm{e}/(\textrm{e}-1)$, obtained by imposing the values of equations \ref{vomax2} and \ref{vomax3} to be equal for $\tau_R / \tau_F = 1 / \textrm{e} \simeq 0.3$.
From equations \ref{rmsout10}, \ref{encapp1} and \ref{vomax3} one obtains the expression for the squared ENC, that is
\begin{equation}
\textrm{ENC}^2 \simeq \left({1+2 \frac{\tau_R}{\tau_F} }  \right)^2
\left(
 i_n^2 \frac{\tau_F - \tau_R }{4 } + e_n^2 C_I^2 \frac{\tau_F - \tau_R }{4 \tau_F \tau_R} + A_f  C_I^2  \ln{ \frac{\tau_F }{\tau_R}}
\right)
 \end{equation}
which to the first order in $\tau_R / \tau_F$ can be also written as
\begin{equation}
\textrm{ENC}^2 \simeq
 i_n^2 \frac{\tau_F + 3 \tau_R }{4} +
 e_n^2 C_I^2 \frac{\tau_F + 3 \tau_R }{4 \tau_F \tau_R}  +
A_f  C_I^2 \frac{\tau_F + 4 \tau_R}{\tau_F} \ln{ \frac{\tau_F }{\tau_R}}
\label{enc1}
\end{equation}
By taking the square root of equation \ref{enc1} we obtain equation \ref{enc2}.

\subsection{Jitter}
\label{jitterapp}

To calculate the impact of the noise of the CSA on the timing resolution of the discriminator, one must consider the overall transfer function of the CSA and of the first stage of the comparator when it is triggering, that is when the voltage at the two inputs is almost equal.
Neglecting hysteresis the transfer function of the first stage of the comparator in the complex frequency domain can be modelled as $ G / (1 + s \tau_C)$.
By combining this with equation \ref{signals1} we obtain the transfer function of the whole chain from the CSA input to the discriminator output, which gives
\begin{equation}
\textrm{TF}_{\textrm{\footnotesize \ TOT}}(s)   \simeq  \frac{1}{s C_F}  \frac{G}{1 + s \tau_C}\frac{s \tau_F}{\left( 1 + s \tau_F \right)}
\label{comps2}
\end{equation}
where equation \ref{signals1} was approximated for $\tau_R \simeq 0$, since bandwidth is now limited by $\tau_C$, which is expected to be larger than $\tau_R$ at least for small values of the input capacitance $C_I$.
As in the case of the squared RMS noise at the output of the CSA alone, which was given by equations \ref{rmsout4}, \ref{rmsout5} and \ref{rmsout6}, we can calculate the squared RMS noise at the output of the first stage of the discriminator. For a current noise source $i_n$ we obtain
\begin{equation}
V_{O i\ \textrm{\footnotesize RMS}}^2 =
\frac{i_n^2}{C_F^2} G^2  \frac{\tau_F^2}{4 ( \tau_F + \tau_C) } 
\simeq
\frac{i_n^2}{C_F^2} G^2  \frac{\tau_C}{8 } 
\label{rmsjitt1}
\end{equation}
for the voltage white noise
\begin{equation}
V_{O e\ \textrm{\footnotesize RMS}}^2 =  
e_n^2 \frac{C_I^2}{C_F^2} G^2   \frac{\tau_F^2}{4 (\tau_F^2 \tau_C + \tau_F \tau_C^2)} 
\simeq
e_n^2 \frac{C_I^2}{C_F^2} G^2   \frac{1}{8 \tau_C} 
\label{rmsjitt2}
\end{equation}
and for the voltage low frequency noise
\begin{equation}
V_{OA\ \textrm{\footnotesize RMS}}^2 =
A_f \frac{C_I^2}{C_F^2}  G^2  \frac{\tau_F^2   }{ \tau_F^2 - \tau_C^2 }  \ln{ \frac{\tau_F }{\tau_C}} 
\simeq
A_f \frac{C_I^2}{C_F^2}  G^2    \frac{1}{2}
\label{rmsjitt3}
\end{equation}
where the expressions were approximated for $\tau_C \simeq \tau_F$.
The sum of these gives the total RMS noise at the output of the first stage of the discriminator. To obtain the corresponding timing resolution, one must divide the voltage noise by the slope of the signals at the output:
\begin{equation}
\sigma_{\textrm{\footnotesize \ Rise}} = \frac{V_{O\ \textrm{\footnotesize RMS}}}{V'_O (t=t_{TH})}
\label{jitterdef1}
\end{equation}
where $t_{TH}$ is the time when the second stage of the discriminator triggers the signal.
By multiplying equation \ref{comps2} by the input charge in excess of threshold, that is $Q-Q_{TH}$, then computing its inverse Laplace transform and differentiating it with respect to time, one obtains
\begin{equation}
V'_O (t) = \frac{Q - Q_{TH}}{C_F} G \frac{ \tau_F}{\tau_F - \tau_C} \left(   \frac{\textrm{e}^{-\frac{t}{\tau_C}}}{\tau_C} - \frac{\textrm{e}^{-\frac{t}{\tau_F}}}{\tau_F} \right)
\label{slopej1}
\end{equation}
which, considering that $\tau_F \simeq \tau_C$, becomes
\begin{equation}
V'_O (t) = \frac{Q - Q_{TH}}{C_F} G \frac{\tau_C - t }{\tau_C^2} \textrm{e}^{-\frac{t}{\tau_C}}
\label{slopej2}
\end{equation}
Assuming that the second stage of the discriminator triggers for $t \ll \tau_C$ equation \ref{slopej2} gives
\begin{equation}
V'_O (t=t_{TH}) = \frac{Q - Q_{TH}}{C_F} \frac{G}{\tau_C}
\label{slopej3}
\end{equation}
By plugging equation \ref{slopej3} together with equations \ref{rmsjitt1}, \ref{rmsjitt2} and \ref{rmsjitt3}  into equation \ref{jitterdef1} one obtains
\begin{equation}
\sigma^2_{\textrm{\footnotesize \ Rise}} \simeq \frac{1}{\left( Q-Q_{TH} \right)^2}
\left(i_n^2  \frac{\tau_C^3}{8} + 
e_n^2 C_I^2 \frac{\tau_C}{8 } +
A_f C_I^2   \frac{\tau_C^2}{2}
 \right)
\label{jittermainj0}
\end{equation}
By taking the square root of equation \ref{jittermainj0} one obtains equation \ref{jittermain1}.

\end{document}